\documentstyle[twocolumn,floats,epsf,ifthen,aps,prb]{revtex} 

\begin{document} 

\preprint{draft} 
 

\title{ 
Output spectrum of a detector measuring quantum oscillations } 

\author{Alexander N. Korotkov}
\address{ 
Department of Physics, State University of New York at Stony Brook, 
Stony Brook, NY 11794-3800
} 
\date{\today} 
 
\maketitle 
 
\begin{abstract} 
        We consider a two-level quantum system (qubit) which is continuously 
measured by a detector and calculate the spectral density of the 
detector output. In the weakly coupled case the spectrum exhibits 
a moderate peak at the frequency of quantum oscillations and 
a Lorentzian-shape increase of the detector noise at low frequency. 
With increasing coupling the spectrum transforms into 
a single Lorentzian corresponding to random jumps between two states. 
        We prove that the Bayesian formalism for the selective evolution
of the density matrix gives the same spectrum as the conventional master 
equation approach, despite the significant difference in interpretation. 
The effects of the detector nonideality and the finite-temperature
environment are also discussed. 
\end{abstract} 
 
\narrowtext 
 

\section{Introduction}

        The long-standing and still controversial problem of 
quantum measurements \cite{Wheeler} is gradually becoming a 
practical issue due to the development of experimental
techniques capable of measurements at the quantum border 
(see, e.g.\ Refs.\ \cite{Aspect,Itano,Braginsky,Buks,Nakamura,Sprinzak}).
The renewed interest in this subject is caused by the needs of 
quantum computing,\cite{Bennett} since the measurement of 
an entangled and possibly evolving quantum state by a 
realistic detector in a realistic environment presents a nontrivial 
problem.  

        Despite the experimental proof \cite{Aspect} of the violation 
of Bell's inequality \cite{Bell} that rejects the idea of hidden
variables, the origin of the randomness of a quantum measurement result 
(the problem known as the ``wavefunction collapse'') remains 
controversial. Important insight into this problem has been provided  
by the development of the theory of continuous quantum measurement, 
which generalizes the ``orthodox'' case \cite{Neumann} of instantaneous 
measurement. There are two different theoretical approaches. 
In the first approach \cite{Zurek} based on the theory of interaction
with a dissipative environment, \cite{Caldeira} the evolution of the
density matrix of the measured system is {\it averaged} over a 
complete {\it ensemble} of 
measurements, thus leading to the deterministic equation. This 
is the best-known approach, at least in the solid-state community, 
and so can be called ``conventional''. The other approach 
\cite{Gisin,Carmichael,Gagen,Plenio,Mensky} (more developed in quantum
optics) studies the random evolution 
of the quantum state during a {\it single} realization of the measurement,
so that this evolution is conditioned on (selected by) the particular 
measurement result. Recently \cite{Kor1,Kor2} the latter approach has been 
introduced into the context of solid-state physics using the new 
derivation based on the simple Bayesian analysis of probabilities. 

        In the present paper this Bayesian formalism is applied 
to the calculation of the spectral density of the detector current 
when a two-level quantum system (qubit) is measured continuously. 
As a particular example, we consider a double-quantum-dot occupied 
by one electron, 
the location of which is measured by a quantum point contact
nearby.\cite{Gurvitz,Kor1} Another example of the setup is 
a single-Cooper-pair box, being measured by a single-electron 
transistor.\cite{Shnirman,Kor2,Makhlin} 
One more possible example is the continuous measurement of two flux 
states of a SQUID by another inductively coupled SQUID.\cite{Han}

 We show that in the 
weak-coupling case when the quantum (Rabi) oscillations of the qubit
state are only slightly perturbed by the detector, the corresponding 
peak in the spectral density of the detector current can be up to 4 times
higher than the noise pedestal (see also Ref.\ \cite{Kor-Av}). 
As the coupling increases,  
the Quantum Zeno effect \cite{Misra} becomes significant leading to the 
Lorentzian shape of the spectral density centered at zero frequency. 

It is important to notice that there should be 
no difference between the predictions of the conventional approach
and the approach of selective evolution unless the measurement
result affects the system evolution (for example, via the feedback
loop). We prove this equivalence explicitly for the detector spectral
density (if there is no feedback).
        Despite the same final result, the interpretations are different:
in the Bayesian formalism a significant contribution to the spectrum comes
from the correlation between the detector noise and the system evolution, 
while this correlation is absent in the conventional approach.
In the paper we also discuss the extension of the Bayesian formalism
to the case of additional weak interaction of the two-level system
with a finite-temperature environment.\cite{Kor-Av} 

\section{Bayesian formalism} 

        For a two-level quantum system described by the 
Hamiltonian 
        \begin{equation}
{\cal H}_0=\frac{\varepsilon }{2}\, (c_1^\dagger c_1-c_2^\dagger c_2) 
+H (c_1^\dagger c_2+c_2^\dagger c_1)
        \end{equation}  
(where $\varepsilon$ is the energy asymmetry and $H$ is the tunneling 
strength) 
the evolution of its density matrix $\rho_{ij}$ due to continuous 
measurement is given in the conventional approach 
\cite{Zurek,Caldeira,Gurvitz,Shnirman,Kor-Av} by the equations 
        \begin{eqnarray}
&&\dot{\rho}_{11}=    -2 \, \frac{H}{\hbar} \,\mbox{Im} 
\rho_{12}, \,\,\,\, \rho_{11}+\rho_{22}=1, 
        \label{conv1}\\
&& {\dot\rho}_{12}=  \imath \, \frac{\varepsilon}{\hbar}\, \rho_{12}+ 
        \imath \, \frac{H}{\hbar }\, 
(\rho_{11}-\rho_{22})
-\Gamma \rho_{12},  
        \label{conv2} \end{eqnarray}
where the dephasing rate $\Gamma$ due to measurement 
depends on the type of the detector used.
\cite{Buks,Gurvitz,Shnirman,Kor-Av,Aleiner,Levinson,Stodolsky} 

\begin{figure}
\vspace{0.1cm} 
\centerline{
\epsfxsize=2.0in 
\hspace{-0.1cm}
\epsfbox{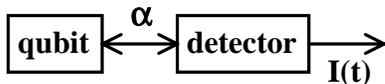}
} 
\vspace{0.4cm}
\caption{Schematic of a qubit continuously measured 
by a detector with output signal $I(t)$. 
 }
\label{schematic}\end{figure}

        Eqs.\ (\ref{conv1})--(\ref{conv2}) describe the averaged evolution.
In contrast, to analyze the single measurement process we need 
the selective (conditional) evolution of $\rho_{ij}$ which in the Bayesian 
formalism is described by equations\cite{Kor1} 
        \begin{eqnarray}
\dot{\rho}_{11}= &&  -2\, \frac{H}{\hbar }\,\mbox{Im} \rho_{12}
 +\frac{2\Delta I}{S_0}\, \rho_{11}\rho_{22} [I(t)-I_0], 
        \label{Bayes1}\\
 {\dot\rho}_{12}= &&  \imath \, \frac{\varepsilon}{\hbar}\,\rho_{12}
        + \imath \, \frac{H}{\hbar } \, (\rho_{11}-\rho_{22}) 
\nonumber \\ 
&&  - \frac{\Delta I}{S_0} \, ( \rho_{11}-  \rho_{22})  
[I(t)-I_0] \, \rho_{12} -\gamma \, \rho_{12},  
        \label{Bayes2} \end{eqnarray}
where $I(t)$ is the detector current, $I_0=(I_1+I_2)/2$, 
$I_1$ and $I_2$ are the average 
currents corresponding to two localized states of the qubit, 
$\Delta I=I_1-I_2$ (notice the different sign in the definition used
in Ref.\ \cite{Kor1}), and $S_0$ is the low-frequency spectral density 
of the detector shot noise (which is assumed to be constant in the 
frequency range of interest and to be practically independent of the qubit 
state).
        The detector nonideality is described by the dephasing contribution 
$\gamma =\Gamma - (\Delta I)^2/4S_0 \geq 0$ due to interaction with 
``pure environment'' (which does not affect the detector current).
In particular, since $\Gamma =(\Delta I)^2/4S_0$ for symmetric quantum
point contact (see Refs.\ \cite{Gurvitz,Buks,Kor-Av,Aleiner}),
it represents an ideal detector,\cite{Kor1}  
$\eta =1$, where $\eta = 1-\gamma /\Gamma$ is the ideality factor.
In contrast, the single-electron transistor in a typical operation point 
is a significantly nonideal detector, $\eta \ll 1$. 
\cite{Kor1,Shnirman,correction} 
(Actually, for the single-electron transistor Eq.\ (\ref{Bayes2})
can be further improved,\cite{Kor2} however, it does not make a difference 
for the purposes of the present paper.) 
        The SQUID can be an ideal detector when its sensitivity 
is quantum-limited. \cite{Danilov,Averin} Since the typical 
output signal from the SQUID is the voltage (not the current), this requires
a minor modification of the formalism; so in this paper we do not consider 
the SQUID case explicitly even though all the results can be easily 
translated into SQUID language. 

        Eqs.\ (\ref{Bayes1})--(\ref{Bayes2}) allow us to calculate
the evolution of $\rho_{ij}$ if the detector output $I(t)$ is known
from the experiment. In order to simulate the measurement we need the
replacement \cite{Kor1} 
        \begin{equation}
I(t)- I_0 \, = \, \Delta I (\rho_{11}-\rho_{22})/2 +\xi (t) , 
        \label{Bayes3}\end{equation}
where the random process $\xi (t)$ has zero average and ``white'' spectral
density $S_\xi = S_0$.

        Notice that in order to consider the detector as a device with 
classical output signal,  Eqs.\ (\ref{Bayes1})--(\ref{Bayes2}) 
essentially use the Markov approximation and assume that 
the typical frequency of the internal detector 
evolution (on the order of $S_0/e^2 \sim I_0/e$) is much higher than 
the typical frequency $\max ( \Omega ,\Gamma )$ of $\rho_{ij}$ evolution 
(here $\Omega \equiv \sqrt{4H^2+\varepsilon^2}/\hbar$ is the frequency of
unperturbed quantum oscillations). In particular, this condition requires 
the detector to be ``weakly responding'', $|\Delta I|\ll I_0$,  
that allows us to use the linear response approximation. 

 Averaging of Eqs.\ (\ref{Bayes1})--(\ref{Bayes2}) 
over all possible measurement results (i.e.\ over random contribution 
$\xi (t)$) reduces them to Eqs.\ (\ref{conv1})--(\ref{conv2}).
Notice that the stochastic equations are written 
in Stratonovich form which preserves the usual calculus rules, while
averaging would be easier in It\^o form.\cite{Oksendal} 

\begin{figure}[t]
\vspace{0.1cm} 
\centerline{
\epsfxsize=3.0in
\hspace{-0.2cm} 
\epsfbox{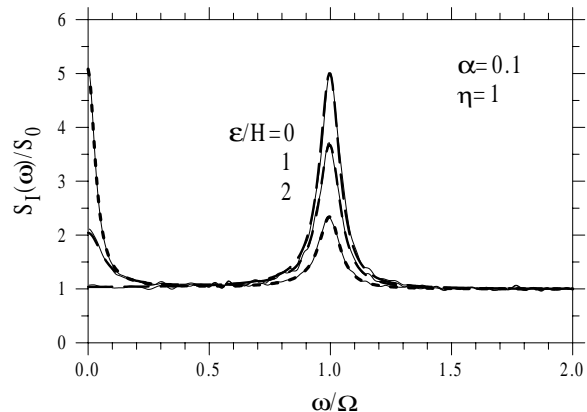}
} 
\vspace{0.4cm}
\caption{Spectral density $S_I(\omega )$ of the detector current 
for weak coupling ($\alpha =0.1$) of an ideal detector ($\eta =1$)  
with a two-level system ($\varepsilon /H =0$, 1, and 2). 
The results of the Monte-Carlo calculations 
using Bayesian formalism are shown by solid lines, while the dashed 
lines (almost coinciding with solid lines) are calculated using 
the master equation approach. 
 }
\label{M-C}\end{figure}

\section{Weak coupling}

\begin{figure}[t]
\vspace{0.1cm}
\centerline{
\epsfxsize=2.7in
\hspace{-0.1cm} 
\epsfbox{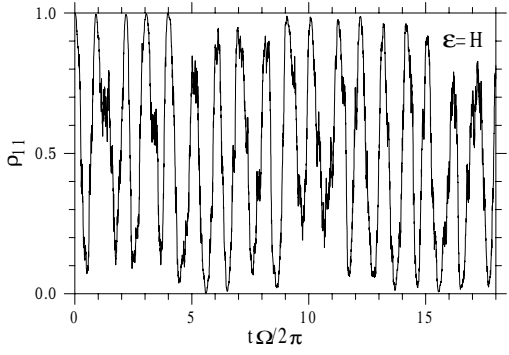}
} 
\vspace{0.4cm}
\caption{ A particular realization of the evoltion of $\rho_{11}(t)$ 
due to continuous measurement for $\varepsilon /H=1$, $\alpha =0.1$ and 
$\eta =1$. Notice the fluctuation of both the phase and the asymmetry
of oscillations. 
 }
\label{asym}\end{figure}

\begin{figure}[ht]
\vspace{0.1cm}
\centerline{
\epsfxsize=3.0in
\hspace{-0.1cm}
\epsfbox{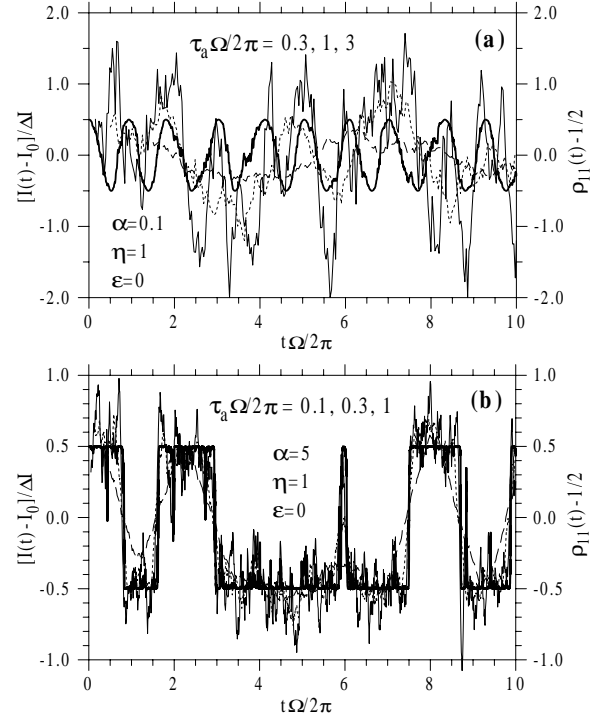}
} 
\vspace{0.4cm}
\caption{A particular realization of $\rho_{11}$ evolution (thick line) 
and the corresponding detector output $I(t)$ (thin solid, dotted, 
and dashed lines) averaged using rectangular 
windows with different 
time constants $\tau_a$. (a) -- weak coupling case, $\alpha =0.1$, 
(b) -- strong coupling case, $\alpha =5$. 
 }
\label{I(t)}\end{figure}

        Using Eqs.\ (\ref{Bayes1})--(\ref{Bayes3}) and the Monte-Carlo 
method (similar to Ref.\ \cite{Kor1}) we can calculate in a straightforward 
way the spectral density $S_I(\omega )$ of the detector current $I(t)$.
Solid lines in Fig.\ \ref{M-C} show the results of such calculations for 
the ideal detector, $\eta =1$, and weak coupling between the qubit  
and the detector, $\alpha =0.1$, where $\alpha \equiv 
\hbar (\Delta I)^2/8S_0 H$ ($\alpha$ is 8 times less than the parameter
${\cal C}$ introduced in Ref.\ \cite{Kor1}). One can see that in 
the symmetric case, $\varepsilon =0$, the peak at the frequency of quantum
oscillations is 4 times higher than the noise pedestal, $S_I(\Omega )
=5S_0$ while the peak width is determined by the coupling strength $\alpha$ 
(see Fig.\ \ref{transition} below). In the asymmetric case, 
$\varepsilon \neq 0$, 
the peak height decreases (Fig.\ \ref{M-C}), while the additional 
Lorentzian-shape increase of $S_I(\omega )$ appears at low frequencies. 
The origin of this low-frequency feature is the slow fluctuation of the 
asymmetry of $\rho_{11}$ oscillations (Fig.\ \ref{asym}). 
In case $\varepsilon =0$ the amplitude of $\rho_{11}$ oscillations 
is maximal (see thick line in Fig.\ \ref{I(t)}a), hence there is no  
such asymmetry and the low-frequency feature is absent, while the spectral 
peak at the frequency of quantum oscillation is maximally high. 

        In order to understand the factor 4 for the maximum peak height,  
let us consider the case 
$\alpha \ll 1$, $\varepsilon =0$, and $\eta =1$. Then the selective 
evolution can be written as the quantum oscillations 
with slowly fluctuating phase $\varphi (t)$:
        \begin{eqnarray} 
        z(t)\equiv && \rho_{11}(t)-\rho_{22}(t)=
\cos \phi (t), \,\, \phi= \Omega t +\varphi (t) ,
        \\
&& \rho_{12}=(\imath /2) \sin \phi (t) 
        \end{eqnarray} 
(the state becomes pure\cite{Kor1} after an initial transient period 
since $\eta =1$, while the real 
part of $\rho_{12}$ decays because of $\varepsilon =0$). 
From Eqs.\ (\ref{Bayes1}) and (\ref{Bayes3}) we obtain the random phase 
dynamics:
        \begin{equation}
\dot{\varphi} =- \sin \phi \, \frac{\Delta I}{S_0} [\cos \phi \, 
\frac{\Delta I}{2} +\xi (t)].
        \label{phi}\end{equation} 
Since $(\Delta I)^2/2S_0 \ll \Omega $, we can neglect the first
term in the square brackets and average the second contribution 
over $\phi$ that leads to the white spectrum of $\dot{\varphi}$: 
$S_{\dot{\varphi}}(\omega )=(\Delta I)^2/2S_0$. Then the correlation
function $K_z (\tau )\equiv \langle z(t) z(t+\tau )\rangle$ can be calculated 
as $K_z(\tau ) = \cos (\Omega \tau ) \langle\cos \Delta \varphi (\tau )  
\rangle /2 = \cos (\Omega \tau ) \exp [ -(\Delta I)^2\tau /8S_0]/2$ 
and the spectral density $S_z(\omega )\equiv 2\int_{-\infty}^\infty
K_z(\tau )\exp (\imath \omega \tau )d\tau$ has a peak 
in the vicinity of the oscillation frequency, $|\omega -\Omega |\ll \Omega$:
        \begin{equation}
S_z(\omega )= \frac{8S_0}{(\Delta I)^{2}}\,
\frac{1}{1+[8S_0(\omega -\Omega)/(\Delta I)^2]^2}.
        \end{equation} 

        The detector current is given by Eq.\ (\ref{Bayes3}), so
its spectral density consists of three contributions: 
        \begin{equation} 
S_I(\omega )=S_0 + \frac{\Delta I^2}{4} S_z (\omega ) 
+ \frac{\Delta I}{2} [S_{\xi z}(\omega ) +S_{z\xi}(\omega )],  
        \label{3contrib}\end{equation} 
where the last contribution originates from the correlation between 
$\rho_{ij}$ evolution and the detector noise $\xi (t)$. To calculate 
the correlation function $K_{\xi z}(\tau )\equiv \langle \xi(t) z(t+\tau) 
\rangle $ for $\tau >0$, we need to take into account the phase 
shift $d\varphi =- \sin \phi \Delta I S_0^{-1} \xi (t)dt$ during even 
infinitesimally small time duration $dt$, since the amplitude of the 
stochastic function $\xi (t)$ grows with the timescale decrease,
$\xi (t)^2 dt=S_0/2$.
        Using trigonometric formulas and linear expansion in 
$d\varphi$ we obtain $ \langle \xi (t) \cos [ \phi (t)+d\phi + 
\Omega \tau + \Delta \varphi (\tau )] \rangle 
= \Delta I S_0^{-1} \langle \xi (t)^2 dt\rangle  \langle 
\sin \phi (t) \sin [\phi (t)+\Omega \tau ]\rangle  \langle \cos \Delta 
\varphi (\tau ) \rangle $ and finally 
$K_{\xi z}(\tau )=  \Delta I \cos (\Omega \tau)\exp  
[ -(\Delta I)^2\tau /8S_0]/4$. After the Fourier transformation one finds 
that the correlation between $\xi (t)$ and $z(t)$ brings exactly
the same contribution to the detector spectral density (see Eq.\
(\ref{3contrib})) as the term due to $z(t)$ evolution, so 
that 
        \begin{equation}
S_I(\omega ) = S_0 +\frac{4S_0}{1+[8S_0(\omega -\Omega)/(\Delta I)^2]^2}\, . 
        \label{weak}\end{equation}
Thus, the peak corresponding to quantum oscillations is 4 times 
higher than the noise background, while its full width at half height is 
equal to $(\Delta I)^2/4S_0=\alpha \Omega$ (the same peak width has been 
calculated in Ref.\ \cite{Hackenbroich}). The integral under the peak, 
        \begin{equation}
\int_0^\infty [S(\omega )-S_0] \frac{d\omega}{2\pi} = \frac{(\Delta I)^2}{4}
\, , 
        \label{integral}\end{equation}
has an obvious relation to the average square of the detector current 
variation due to oscillations in the measured system. 
Notice, however, that this integral is twice as large as one would 
expect from the classical harmonic signal, since one half of the peak 
height comes from nonclassical correlation between the qubit evolution
and the detector noise. Classically, Eq.\ (\ref{integral}) would be 
easily understood if the signal was not harmonic but rectangular-like, 
which is obviously not the case.  
Actually, the detector current shows neither clear harmonic 
nor rectangular signal distinguishable from the intrinsic noise 
contribution. Figure \ref{I(t)}a shows the simulation 
of $\rho_{11}$ evolution (thick line) together with the detector 
current $I(t)$. Since $I(t)$ contains white noise, it necessarily 
requires some averaging. Thin solid, dotted, and dashed lines show 
the detector current averaged with different time constants $\tau_a$: 
$\tau_a \Omega/ 2\pi =0.3$, 1, and 3, respectively. For weak averaging 
the signal  
is too noisy, while for strong averaging individual oscillation periods 
cannot be resolved either, so quantum oscillations can never be
observed {\it directly} by a continuous measurement (although  they 
can be {\it calculated} using Eqs.\ (\ref{Bayes1})--(\ref{Bayes2})).
This unobservability is revealed in the relatively low peak height of the 
spectral density of the detector current.

\section{Arbitrary coupling}  

\begin{figure}[ht]
\vspace{0.1cm}
\centerline{
\epsfxsize=3.0in
\hspace{-0.2cm}
\epsfbox{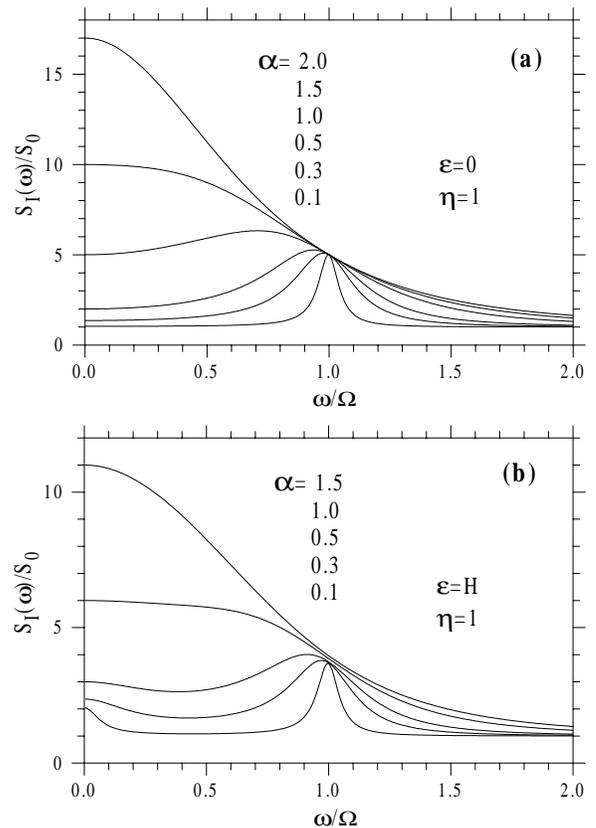}
} 
\vspace{0.4cm}
\caption{The detector current spectral density $S_I(\omega )$ for
$\eta =1$ and different coupling $\alpha$ with (a) symmetric  
($\varepsilon =0$) and (b) asymmetric ($\varepsilon /H=1$) qubit. 
 }
\label{transition}\end{figure}

        The situation changes as the coupling between the detector 
and qubit increases, $\alpha \gtrsim 1$. 
The strong influence of measurement destroys quantum oscillations,  
and the Quantum Zeno effect \cite{Misra} develops, 
so that for $\alpha \gg 1$ the qubit performs random jumps between 
two localized states (see Fig.\ \ref{I(t)}b). In this case 
the properly averaged detector current follows pretty well the 
evolution of the qubit (however, the unsuccessful tunneling ``attempts'' 
still cannot be directly resolved), and the spectral density of $I(t)$ can 
be calculated using the classical theory of telegraph 
noise\cite{Machlup} leading to the Lorentzian shape of $S_I(\omega )$. 
Figure \ref{transition}a shows the gradual transformation of the 
spectral density with the increase of the coupling $\alpha$ for
a symmetric qubit, $\varepsilon =0$, and an ideal detector, $\eta =1$. 
The results for an asymmetric qubit, $\varepsilon /H=1$, are shown
in Fig.\ \ref{transition}b.

        The curves in Fig.\ \ref{transition} as well as the dashed
curves in Fig.\ \ref{M-C} are calculated using the conventional
master equation approach which gives the same results for the
detector spectral density as the Bayesian formalism 
(we will prove this later). In the 
conventional approach we should assume no correlation between the
detector noise and the qubit evolution (the last term in Eq.\
(\ref{3contrib}) is absent) while the correlation function 
$K_{\hat z}(\tau )$ should be calculated considering $z(t)$ not as an 
ordinary function but as an operator. Then the calculation of 
$\langle \hat z(t) \hat z(t+\tau )\rangle $ can be essentially 
interpreted as follows. The first (in time) operator $\hat z(t)$ 
collapses the qubit into one of two eigenstates which correspond
to localized states, then during time $\tau$ the qubit performs 
the evolution described by conventional Eqs.\ (\ref{conv1})--(\ref{conv2}), 
and finally the second operator $\hat z(t+\tau )$ gives the
probability for the qubit to be measured in one of two localized
states. (Of course, this procedure can be done purely formally,\cite{Kor-Av} 
without any interpretation.) Notice that there is complete symmetry 
between states ``1'' and ``2'' even for $\varepsilon \neq 0$ 
(in particular, in the stationary state 
$\rho_{11}=\rho_{22}=1/2$), so the evolution after the first collapse 
can be started from any localized state leading to the same contribution 
to the correlation function. In this way we obviously get 
$K_{\hat z}(\tau )=\rho_{11}(\tau )- 
\rho_{22}(\tau )$ where $\rho_{ii}$ is the solution of Eqs.\
(\ref{conv1})--(\ref{conv2}) with the initial conditions $\rho_{11}(0)=1$
and $\rho_{12}(0)=0$. 

        For the symmetric qubit, $\varepsilon =0$, these equations 
can be easily solved analytically, and finally we obtain:
        \begin{equation}
S_I(\omega ) = S_0 + \frac{\Omega^2 (\Delta I)^2 \Gamma }
        {(\omega^2-\Omega^2)^2+\Gamma^2\omega^2} \, , 
       \label{sym} \end{equation}
where $\Gamma =\eta^{-1} (\Delta I)^2/4S_0 = \alpha \eta^{-1}\Omega$. 
This equation obviously transforms into Eq.\ (\ref{weak}) for $\eta =1$ 
and $\alpha \ll 1$. Notice that for weak coupling with a nonideal detector, 
$\eta < 1$ and $\alpha \eta^{-1} \ll 1$,  
the peak height of $S_I(\omega )$ is equal to $4\eta S_0$,  
while the linewidth $\alpha \eta^{-1}\Omega $ of the peak is 
$\eta^{-1}$ times wider than for the ideal detector. 
As the coupling increases, the linewidth grows and the oscillation 
frequency decreases: \cite{Hackenbroich} 
 $\omega_{osc} = \Omega [1- (\alpha /2\eta )^2)^{1/2}] $. 
The transition into the overdamped regime occurs at $\alpha \eta^{-1}> 2$ 
while the peak-like feature disappears at $\alpha \eta^{-1} >\sqrt{2}$.
For $\alpha \eta^{-1} >2$ the spectral density consists of two 
Lorentzians [$\omega_{1,2}=\Gamma /2 \mp (\Gamma^2/4 - \Omega^2)^{1/2}$]  
centered at zero frequency, 
with the negative sign and the smaller amplitude $A_2$ of the
second Lorentzian which has higher cutoff frequency, 
$A_2/A_1=-\omega_1/\omega_2$.  In the case $\alpha \eta^{-1} \gg 1$,  
which corresponds to the well developed Quantum Zeno effect, 
$S_I(\omega ) -S_0$ has purely Lorentzian shape $(\Delta I)^2\omega_1
/(\omega^2+\omega_1^2)$ with 
$\omega_1 =\Omega^2/\Gamma = \Omega \eta/ \alpha $. 

        For the asymmetric qubit, $\varepsilon \neq 0$, the spectral 
density can in principle also be calculated analytically but the expressions 
would be too lengthy and it is simpler to use numerical solution 
of Eqs.\ (\ref{conv1})--(\ref{conv2}). The analytical formula for
the weak coupling limit is 
        \begin{eqnarray}
S_I(\omega )  = && S_0 + \frac{\eta S_0 \varepsilon^2/H^2}
{1+(\omega\hbar^2\Omega^2/4H^2\Gamma )^2}
        \nonumber \\ 
&& +\frac{4\eta S_0 (1+\varepsilon^2/2H^2)^{-1}} 
{1+[(\omega -\Omega)/\Gamma(1-2H^2/\hbar^2\Omega^2)]^2} \, .
        \label{Sasym}\end{eqnarray}
The spectral peak and the low-frequency Lorentzian become wider with 
the coupling increase since $\Gamma = \alpha \eta^{-1}\Omega$, and  
for $|\varepsilon /H| <1/\sqrt{2}$ the overdamped regime starts from 
$\Gamma =\Gamma_1$ where $\Gamma_{1,2}^2=(\Omega^2/2a)[b\mp (b^2 -4a)^{1/2}]$,
$b\equiv 1/4-27a^2/4+9a/2$, $a\equiv \varepsilon^2 /(4H^2+\varepsilon^2)$. 
At $\Gamma >\Gamma_2$ the dynamics formally returns to the underdamped 
regime, however, the peak linewidth is much larger than the frequency 
and so $S_I(\omega )$ is monotonous. For $|\varepsilon /H| >1/\sqrt{2}$ 
the overdamped regime does not occur. 
        In both cases in the limit of large $\Gamma$ the spectral density
has almost Lorentzian shape with the cutoff frequency 
$\omega_1=4H^2/\hbar^2\Gamma $.

        One can check that the spectral densities given by Eqs.\ 
(\ref{sym}) and (\ref{Sasym}) satisfy the integral condition 
(\ref{integral}), 
which remains valid for arbitrary parameters $\alpha$, $\varepsilon /H$, 
and $\eta$, because of the equation $K_I(+0)=(\Delta I/2)^2$.

\section{Equivalence of two approaches} 

Comparing two derivations of $S_I(\omega )$ in the case $\alpha \ll 1$, 
$\eta =1$, and $\varepsilon =0$, 
we see that $K_{\hat z}(\tau )$ is twice larger than $K_z(\tau )$ 
because in the conventional approach the evolution always starts 
from the localized state while  
in the Bayesian approach it starts from arbitrary phase of quantum 
oscillations. This difference exactly compensates for the absence of the 
last term in Eq.\ (\ref{3contrib}) in the conventional approach. 

        Let us prove explicitly that the two approaches give the 
same result for $S_I(\omega )$ in a general case. In order to calculate 
$K_{\xi z}(\tau )$ for $\tau>0$ using the Bayesian formalism, 
let us first average the product $\xi(t_0)z(t_0+\tau )$ over random 
$\xi (t)$ during 
time period $t_0<t<t_0+\tau $, fixing the same conditions at $t=t_0$.
Then we can use conventional Eqs.\ (\ref{conv1})--(\ref{conv2})
[regarded as Eqs.\ (\ref{Bayes1})--(\ref{Bayes3}) averaged over random 
$\xi (t)$] with the initial condition 
$\rho_{ij}(t_0+0)= \rho_{ij}(t_0)+d\rho_{ij}$ where 
        \begin{eqnarray}
&& dz=  \Delta I\, S_0^{-1} [1-z^2(t_0)] \, \xi(t_0)\, dt , 
        \\ 
&& d\rho_{12}= - \Delta I\, S_0^{-1} z(t_0)\, \rho_{12}(t_0)\, 
        \xi(t_0) \, dt  
        \end{eqnarray}
(for simplicity we will refer to 
$z\equiv \rho_{11}-\rho_{22}$ as a component of $\rho_{ij}$). 
Since the sign of $\xi(t_0)$ is arbitrary and averaged  
evolution equations are linear, we need only the fluctuating contribution 
to $\rho_{ij}(t_0+0)$ and, hence, can formally assume that the 
evolution starts from $\rho_{ij}(t_0+0)=d\rho_{ij}$ 
(notice that we can forget the condition 
$\rho_{11}+ \rho_{22}=1$ and use only $z$ and $\rho_{12}$).
Using the relation $\xi(t_0)^2dt=S_0/2$ and the evolution linearity, 
we can formally write 
        \begin{equation}
K_{\xi z}(\tau )=(\Delta I/2) \langle \tilde z(t_0+\tau ) \rangle ,
        \end{equation}
where $\tilde \rho_{ij}$ satisfies Eqs.\ (\ref{conv1})--(\ref{conv2})
with $\tilde z(t_0) =1-z(t_0)^2 $ and $\tilde \rho_{12} (t_0)=-z(t_0)
\rho_{12}(t_0)$, and the averaging over the initial conditions at $t=t_0$ 
should still be done later. Before that let us do a similar formal trick 
for $K_z(\tau )$ representing it as $\langle\tilde z(t_0+\tau )\rangle$ where
the evolution starts from $\tilde z(t_0)=z(t_0)^2$ and $\tilde \rho_{12}
(t_0)= z(t_0)\rho_{12}(t_0)$. Now combining two terms in the detector 
current 
correlation function $K_I(\tau ) \equiv \langle I(t_0) I(t_0+\tau )\rangle 
-\langle I\rangle^2 = (\Delta I/2)^2 K_z(\tau ) + 
(\Delta I/2)K_{\xi z}(\tau )$ (here $\tau >0$), we see that it can be 
written as $(\Delta I/2)^2\langle \tilde z(t_0+\tau )\rangle$ where 
$\tilde z(t_0)=1$ and $\tilde \rho_{12}(t_0)=0$. Thus we have exactly 
arrived at the expression of the conventional approach, in which
the evolution always starts from the localized state, regardless 
of the actual quantum state at $t=t_0$. This proof is obviously valid
for arbitrary $\alpha$, $\eta$, and $\varepsilon /H$.
Despite the same result in two approaches, the interpretations are
quite different since the Bayesian approach allows us to follow
the qubit evolution during the measurement process, while the
conventional approach gives only the average characteristics.

\section{Finite-temperature environment}

        In this section we will discuss how to introduce
the finite-temperature environment into Eqs.\ (\ref{Bayes1})--(\ref{Bayes3})
of the Bayesian formalism. Notice that so far there has been complete
symmetry between the states ``1'' and ``2'' even for finite energy difference 
$\varepsilon$, while the finite temperature effects would be expected to 
lead to different average populations of these states. 
Such  symmetry requires an implicit assumption that the typical energy 
in the detector (voltage or 
temperature) is much larger than the energies involved in the qubit dynamics. 
So, the absence of temperature in the formalism does not mean that 
it is zero or very large, it just means that the temperature effects
are not important. 
Now let us assume that besides the detector, the qubit is coupled
to an additional finite-temperature environment, which creates
an asymmetry between states ``1'' and ``2'' when $\varepsilon \neq 0$.

        While the case of finite coupling of a two-level system 
with an environment presents a really difficult problem,\cite{Caldeira} 
the weak coupling limit can be treated in a simple way. In the standard 
method \cite{weakcoupling} it is described by the equations 
        \begin{eqnarray}
&& \dot \rho_{++}=-\gamma_1 (\rho_{++}-p_{st}) , \,\,\,\,
        \rho_{++}+\rho_{--}=1 ,
        \label{env1} \\
&& \dot \rho_{+-}=\imath \Omega \rho_{+-} -\gamma_2 \rho_{+-}, 
        \label{env2}\end{eqnarray}
which are written in the diagonal basis (``+'' corresponds to the ground 
state). The temperature $T$ determines the stationary occupation
$p_{st}=[1+\exp (-\hbar\Omega /T)]^{-1}$ of the ground state,  
and the relaxation rates obey inequality \cite{weakcoupling}
$\gamma_1/2 \leq \gamma_2  \ll \Omega$. 

If the coupling of the qubit with the detector is also
weak, $\alpha \eta^{-1} \ll 1$, the evolution due to extra 
finite-temperature environment can be simply added to the evolution 
due to measurement. For this purpose Eqs.\ (\ref{env1})--(\ref{env2})
should be translated into the basis of localized states, so the terms
        \begin{eqnarray}
&& -(A^2\gamma_1 +B^2\gamma_2)(\rho_{11}-1/2) -\gamma_1 A
        (1/2-p_{st})
        \nonumber \\
&& \hspace{0.3cm} +AB(\gamma_1-\gamma_2) 
        \, \mbox{Re} \rho_{12} ,
        \label{add1} \\ 
&& \mbox{where} \,\, A \equiv \varepsilon/\hbar \Omega , \,\,
        B\equiv 2H/ \hbar \Omega ,
        \nonumber \end{eqnarray} 
should be added into Eq.\ (\ref{Bayes1}) for $\dot \rho_{11}$ 
and the terms 
        \begin{eqnarray}
&& -(A^2\gamma_2+B^2\gamma_1)\, \mbox{Re}\rho_{12} +AB(\rho_{11}-
\rho_{22})(\gamma_1-\gamma_2)/2 
        \nonumber \\
&& \hspace{0.3cm} +\gamma_1B(1/2-p_{st}) -\imath \gamma_2\, 
\mbox{Im}\rho_{12}  
        \label{add2} \end{eqnarray}
should be added into Eq.\ (\ref{Bayes2}) for $\dot \rho_{12}$.
The same terms should obviously be added into Eqs.\ 
(\ref{conv1})--(\ref{conv2}) for the conventional approach. 
(Of course, this generalization is purely phenomenological and is 
limited to the weak coupling regime, so the effect of
Eqs.\ (\ref{add1})--(\ref{add2}) can be considered only at the 
timescale longer than oscillation period.) 

        In the generalized case it is still possible to prove that
the results of the Bayesian formalism for the detector current 
spectral density $S_I(\omega )$ exactly coincide with the results
of the conventional approach. The essential difference from the 
proof above is nonzero stationary solution ($z_{st},\, \rho_{12,st}$)  
of modified Eqs.\ (\ref{conv1})--(\ref{conv2}) when $p_{st}\neq 1/2$. 
It is convenient to consider homogeneous evolution equations (with
$p_{st}=1/2$) simply shifting $z(t)$ and $\rho_{12}(t)$ by the
stationary values. Using the same idea as in the proof above we
can show that in the Bayesian approach $K_I(\tau )$ for $\tau >0$ 
can be written as $(\Delta I/2)^2 \langle \tilde z (t_0+\tau ) \rangle$ 
where $\tilde \rho_{ij}$ satisfies homogeneous modified Eqs.\
(\ref{conv1})--(\ref{conv2}) with $\tilde z(t_0)=1 -2z_{st} z(t_0)
+z_{st}^2$ and $\tilde \rho_{12}(t_0)= -z(t_0)\rho_{12,st} -
z_{st}\rho_{12}(t_0)+z_{st}\rho_{12,st}$. After the averaging
over initial states these initial conditions can be replaced
with $\tilde z(t_0)=1-z_{st}^2$ and $\tilde \rho_{12} (t_0) = 
-z_{st}\rho_{12,st}$. 

        Now let us show that we get the same $K_I(\tau )$ in the 
conventional approach.
With the probability $(1+z_{st})/2$ the first operator $\hat z(t_0)$
localizes the qubit into the state ``1''. Then the initial state
for the homogeneous equations is $\tilde z(t_0)=1-z_{st}$, 
$\tilde \rho_{12}(t_0)=-\rho_{12,st}$. With the probability $(1-z_{st})/2$
the evolution starts from the state ``2'', i.e.\ $\tilde z(t_0)=-1-z_{st}$,
$\tilde \rho_{12} (t_0)=-\rho_{12,st}$. Adding two contributions with 
opposite signs we see again that for $\tau >0$, $K_I(\tau )= 
(\Delta I/2)^2 \tilde z (t_0+\tau ) $ where $\tilde z$ can be found
as a solution of Eqs.\ (\ref{conv1})--(\ref{conv2}) modified by
Eqs.\ (\ref{add1})--(\ref{add2}) without inhomogeneous terms, 
with the initial condition $\tilde z(t_0) = 1-z_{st}^2$ and 
$\tilde \rho_{12} (t_0) = -z_{st}\rho_{12,st}$. 
Thus, the correlation function $K_I(\tau )$ and, hence, the spectral 
density $S_I(\omega )$ coincide in the two approaches. 

	It is technically simpler to consider the averaged evolution 
in the diagonal basis rather than in the basis of localized states [for this 
purpose we need to translate the term $-\Gamma \rho_{12}$ from 
Eq.\ (\ref{conv2}) into the diagonal basis and add it into Eqs.\ 
(\ref{env1})--(\ref{env2})]. So, to calculate the correlation function
$K_{\hat z}(\tau )$ analytically, we start the evolution from one of 
the localized states, then consider the averaged evolution in the
diagonal basis (neglecting the rapidly oscillating terms due to measurement),
and make the second projection onto localized states at $t=\tau$. 
Finally we obtain the result \cite{Kor-Av} 
	\begin{eqnarray}
S_I(\omega ) = &&  S_0 + \frac{(\Delta I)^2}{W_t}\, 
\left [ \frac{\varepsilon^2}{\hbar^2\Omega^2} -z_{st}^2 \right] 
 \frac{1}{1+(w/W_t)^2} 
	\nonumber \\
&& 	+ \frac{2(\Delta I)^2 H^2}{W_0 \hbar^2 \Omega^2} \,
	\frac{1}{1+[(\omega-\Omega )/W_0]^2} 
	\label{Senv}\end{eqnarray}
where 
	\begin{eqnarray}
&& z_{st}= \frac{\varepsilon }{\hbar\Omega }\,
\frac{1}{1+4H^2\Gamma /\gamma_1\hbar^2\Omega^2}
 \, \tanh (\frac{\hbar \Omega }{2T}),    
	\\ 
&& W_t = \gamma_1 +\frac{4\Gamma H^2}{\hbar^2\Omega^2 },
	\\
&& W_0 = \gamma_2 +\frac{\Gamma}{2} 
(1+\frac{\varepsilon^2}{\hbar^2\Omega^2}) .
	\end{eqnarray}

        Let us emphasize that the effect of finite-temperature 
environment is not generally equivalent to the nonideality of the detector 
described by finite $\gamma$ in Eq.\ (\ref{Bayes2}). 
	As an example, in the case of extra environment 
the right hand part of Eq.\ (\ref{integral}) should be multiplied
by the factor $1-z_{st}^2$ which disappears ($z_{st}=0$) only 
if $T=\infty$ or $\varepsilon =0$. 

	Comparing Eqs.\ (\ref{Senv}) and (\ref{Sasym}) one can see
that within the accuracy of weak coupling approximation  
the change of $S_I(\omega )$ due to
extra environment can be reduced to the detector 
nonideality, $\eta <1$, in two cases. 
If $|\varepsilon /H| \ll 1$, then Eqs.\ (\ref{Senv}) and (\ref{Sasym})
coincide at arbitrary temperature $T$ for 
$\eta = (1+2\gamma_2/\alpha\Omega )^{-1}$.  
For asymmetric qubit, 
$|\varepsilon /H| \gtrsim 1$, the equivalence is possible only 
at high temperatures, $T \gg \hbar\Omega$, and requires 
conditions 
	\begin{eqnarray}
&& \gamma_2=\gamma_1(1+\varepsilon^2/2H^2)/2 ,
	\label{enveqv1}	\\
&& \eta^{-1} = 1+(1+\varepsilon^2/4H^2)^{3/2}\gamma_1/\alpha\Omega  .
	\label{enveqv2}\end{eqnarray}

\begin{figure}
\vspace{0.1cm}
\centerline{
\epsfxsize=3.0in
\hspace{-0.2cm} 
\epsfbox{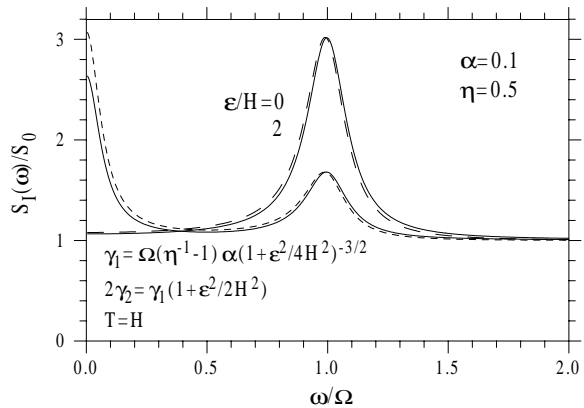}
} 
\vspace{0.4cm}
\caption{ Spectral density $S_I(\omega )$ for 
the nonideal detector (dashed lines)  
and for the case of the ideal detector and weak extra coupling with 
the finite-temperature environment, $T=H$ (solid lines). 
 }
\label{envir}\end{figure}

        Figure \ref{envir} shows the numerically calculated 
spectral density $S_I(\omega )$ 
of the detector current for a nonideal detector, $\eta =0.5$ (dashed 
lines) and for an ideal detector but extra coupling of the qubit 
to the environment at temperature $T=H$ (solid lines). The 
rates $\gamma_1$ and $\gamma_2$ 
are chosen according to Eqs.\ (\ref{enveqv1}) and (\ref{enveqv2}). 
For the symmetric qubit, $\varepsilon =0$, the results of two models
practically coincide. In contrast, the solid and dashed lines for 
$\varepsilon =2H$ significantly differ from each other at low frequencies, 
while the spectral peak at $\omega \sim \Omega$ is fitted quite well.

\section{Conclusion} 

        Using both Bayesian and conventional approaches, we have 
calculated the spectral density $S_I(\omega )$ of the 
detector current when a two-level quantum system (qubit) is measured 
continuously. Depending on the coupling strength, there is a gradual 
transition from quantum oscillations to quantum jumps. 
This results in a transition from the peak-like spectral density
to the Lorentzian shape of $S_I(\omega )$. The maximal height of the
peak at the frequency of quantum oscillation is shown to be 4 times 
the shot noise pedestal (see also Ref.\ \cite{Kor-Av}). 

In the simple case of weak coupling between a symmetric qubit and an ideal
detector, the height of the spectral peak is twice 
as high as the classical result for a harmonic signal. In the Bayesian
approach this is explained by the significant correlation between
the detector noise and the further evolution of the measured system 
due to quantum back-action. In contrast, in the conventional approach this 
fact is the consequence of discrete eigenvalues of $\hat z$ operator,  
which corresponds to the magnitude measured by the detector.
(In other words, this operator ``collapses'' the system into one
of two eigenstates, and that is why the averaged product of two 
operators is twice as large as that for a classical harmonic signal.) 
So, even though the results for $S_I(\omega )$ coincide in two approaches, 
the interpretations are significantly different since the ``abrupt'' 
collapse is replaced in the Bayesian approach by the ``continuous'' 
collapse related to the noisy detector output. 

	It is important to notice that the Bayesian formalism allows us
to monitor continuously the {\it phase} of quantum oscillations. This
makes it possible to tune the phase using the feedback control of 
the qubit parameters. If the real-time calculations using Eqs.\
(\ref{Bayes1})--(\ref{Bayes2}) and  fast feedback loop were available
in an experiment (the typical bandwidth should be larger than $\Gamma$)
then the random diffusion of the oscillation phase could be eliminated
and the qubit could ``stay fresh'' for a very long time. The suppression
of qubit dephasing using the feedback control of the tunneling
strength $H$ has been confirmed by the Monte-Carlo simulations. The
elimination of the phase diffusion gives rise to the $\delta$-function peak
in the detector spectral density $S_I(\omega )$ at the frequency $\Omega$.
A detailed analysis
of this situation is beyond the scope of the present paper.

        The author thanks D. V. Averin, J. E. Lukens, and K. K. Likharev 
for fruitful discussions. 
 The work was partly supported by AFOSR.


\begin{references}


\bibitem{Wheeler} {\it Quantum Theory of Measurement}, ed. by
        J. A. Wheeler and W. H. Zurek (Princeton Univ. Press, 
        1983). 

\bibitem{Aspect} A. Aspect, J. Dalibar, and G Roger, Phys. Rev. Lett.
        {\bf 49}, 1804 (1982). 

\bibitem{Itano} W. M. Itano, D. J. Heinzen, J. J. Bollinger,
        and D. J. Wineland, Phys. Rev. A {\bf 41}, 2295 (1990).  

\bibitem{Braginsky} V. B. Braginsky and F. Ya. Khalili,
        {\it Quantum measurement} (Cambridge Univ. Press, 1992).

\bibitem{Buks} E. Buks, R. Schuster, M. Heiblum, D. Mahalu,
        and V. Umansky, Nature {\bf 391}, 871 (1998).

\bibitem{Nakamura} Y. Nakamura, Yu. A Pashkin, and J. S. Tsai,
        Nature {\bf 398}, 786 (1999). 

\bibitem{Sprinzak} D. Sprinzak, E. Buks, M. Heiblum, and H. Shtrikman,
        cond-mat/9907162.

\bibitem{Bennett} C. Bennett, Phys. Today, Oct. 1995, 24 (1995).

\bibitem{Bell} J. S. Bell, Physics {\bf 1}, 195 (1964).

\bibitem{Neumann} J. von Neumann, {\it Mathematical Foundations of
        Quantum Mechanics} (Princeton Univ. Press, Princeton, 1955). 

\bibitem{Zurek} W. H. Zurek, Phys. Today, {\bf 44} (10), 36 (1991).

\bibitem{Caldeira} A. O. Caldeira and A. J. Leggett, Ann. Phys. (N.Y.)
        {\bf 149}, 374 (1983). 

\bibitem{Gisin} N. Gisin, Phys. Rev. Lett. {\bf 52}, 1657 (1984).

\bibitem{Carmichael} H. J. Carmichael, {\it An open system approach
        to quantum optics}, Lecture notes in physics (Springer, Berlin, 
        1993).

\bibitem{Gagen} M. J. Gagen, H. M. Wiseman, and G. J. Milburn,
        Phys. Rev. A {\bf 48}, 132 (1993).

\bibitem{Plenio} M. B. Plenio and P. L. Knight, Rev. Mod. Phys. 
        {\bf 70}, 101 (1998).

\bibitem{Mensky} M. B. Mensky, Phys. Usp. {\bf 41}, 923 (1998).

\bibitem{Kor1} A. N. Korotkov, Phys. Rev. B {\bf 60}, 5737 (1999);
        quant-ph/9808026.  

\bibitem{Kor2} A. N. Korotkov, in {\it Proceedings of LT'22} 
        (Helsinki, 1999) and cond-mat/9906439.  

\bibitem{Gurvitz} S. A. Gurvitz, Phys. Rev. B {\bf 56}, 15215 (1997);
        quant-ph/9808058. 

\bibitem{Shnirman} A. Shnirman and G. Sch\"on, Phys. Rev B {\bf 57}, 
        15400 (1998).

\bibitem{Makhlin} Y. Makhlin, G. Sch\"on, and A. Shnirman, 
        cond-mat/0001423. 

\bibitem{Han} S. Han, J. Lapointe, and J. E. Lukens, Phys. Rev. Lett.
        {\bf 66}, 810 (1991). 

\bibitem{Kor-Av} A. N. Korotkov and D. A. Averin, cond-mat/0002203. 

\bibitem{Misra} B. Misra and E. C. G. Sudarshan, J. Math. Phys.
        {\bf 18}, 756 (1977).

\bibitem{Aleiner} I. L. Aleiner, N. S. Wingreen, and Y. Meir,
        Phys. Rev. Lett. {\bf 79}, 3740 (1997).

\bibitem{Levinson} Y. Levinson, Europhys. Lett. {\bf 39}, 299 (1997).

\bibitem{Stodolsky} L. Stodolsky, Phys. Lett. B {\bf 459}, 193 (1999). 

\bibitem{correction} From the corrected Eqs.\ (34) and (37) of 
        Ref.\ \protect\cite{Shnirman} in weakly responding case
	($|\Delta I|\ll I_0$), the ``measurement rate'' is  
        $(\Delta I)^2/4S_0=dE^2(\Gamma_R^2/R_L-\Gamma_L^2/R_R)^2 
      [8e^4\Gamma_L\Gamma_R(\Gamma_L+\Gamma_R)(\Gamma_L^2+\Gamma_R^2)]^{-1}$,
        while the dephasing rate is 
        $\Gamma = dE^2\Gamma_L\Gamma_R/\hbar^2(\Gamma_L+\Gamma_R)^3$,
        that gives the ideality factor  
        $\eta = \hbar^2 (\Gamma_R^2/R_L-\Gamma_L^2/R_R)^2
        (\Gamma_L+\Gamma_R)^2 
        [8e^4\Gamma_L^2\Gamma_R^2(\Gamma_L^2+\Gamma_R^2)]^{-1}$,  
        where $\Gamma_L$ and $\Gamma_R$ are the tunneling rates through
        two junctions of single-electron transistor, $R_{L,R}\gg \hbar/e^2$
        are their tunnel resistances, and $dE$ is the energy coupling
        with the measured system. In the case $\Gamma_L \sim \Gamma_R$ 
	we have significantly nonideal detector, $\eta \ll 1$; however, 
	for $\Gamma_L \ll \Gamma_R$ one obtains $\eta = (\hbar /e^2R_L)^2
	(\Gamma_R /\Gamma_L)^2/8$ that becomes comparable to unity when
	$\Gamma_L/\Gamma_R \lesssim \hbar/e^2R_L$, i.e.\ when the 
	cotunneling contribution becomes important.

\bibitem{Danilov} V. V. Danilov, K. K. Likharev, and A. B. Zorin,
        IEEE Trans. Magn. MAG-{\bf 19}, 572 (1983). 

\bibitem{Averin} D. V. Averin, {\it International symposium on 
	mesoscopic superconductivity} (Atsugi, Japan, 2000), Abstracts,
	 p. 89. 

\bibitem{Oksendal} B. {\O}ksendal, {\it Stochastic differential
        equations} (Springer, Berlin, 1992).

\bibitem{Hackenbroich} G. Hackenbroich, B. Rosenow, and 
        H. A. Weidenm\"uller, Phys. Rev. Lett. {\bf 81}, 5896 (1998). 

\bibitem{Machlup} S. Machlup, J. Appl. Phys. {\bf 25}, 341 (1954). 

\bibitem{weakcoupling} U. Weiss, {\it Quantum dissipative systems} 
        (World Scientific, Singapore, 1993). 
 
\end{references}
\end{document}